\documentclass[prl, reprint, amsmath,amssymb,nofootinbib,floatfix, aps,longbibliography, preprintnumbers]{revtex4-2}
\usepackage{graphicx}
\usepackage{dcolumn}
\usepackage{bbold}
\usepackage{mathtools}
\usepackage[colorlinks=true,urlcolor=myblue,anchorcolor=black,citecolor=myblue,filecolor=black,linkcolor=black,menucolor=blue,linktocpage=true]{hyperref} % should be commented out if the tex file will be compiled with latex in arXiv!!! (pdflatex is fine) 
\usepackage{booktabs}
\usepackage{colortbl}
\usepackage{collcell}
\usepackage{enumerate}
\usepackage{float} % Incompatible with revtex4-2, removed to fix compile error
\usepackage{verbatim}
\usepackage{multirow}
\usepackage{xparse}
\usepackage{slashed}
\usepackage{tabularx}
\usepackage{amsmath}
\usepackage{relsize}
\usepackage{braket}
\usepackage{xcolor}
\usepackage[compat=1.1.0]{tikz-feynman}
\usepackage{contour}%FeynamnDiagrams e contorni

\newcommand{\virg}[1]{``#1''}

\definecolor{myred}{cmyk}{0,0.8,0.5,0.5}
\definecolor{myblue}{cmyk}{0.8, 0.4, 0, 0.2}
\definecolor{mygreen}{rgb}{0.27, 0.64, 0.48}
\definecolor{mygray}{gray}{.95}

\def\max{{\mathrm{max}}}
\def\min{{\mathrm{min}}}

\def\DIS{{\mathsmaller{\mathrm{DIS}}}}

\def\MDM{m_{\mathsmaller{\rm DM}}}
\def\DM{\mathsmaller{\rm DM}}
\def\BH{\mathsmaller{\rm BH}}
\def\LOS{\mathsmaller{\rm LOS}}
\def\GS{\mathsmaller{\rm GS}}

\begin{document}

\title{Did IceCube discover Dark Matter around Blazars?}

\author{Andrea Giovanni De Marchi}
\email{andreagiovanni.demarchi@unibo.it}
\affiliation{Dipartimento di Fisica e Astronomia, Università di Bologna, via Irnerio 46, 40126, Bologna, Italy;}
 \affiliation{INFN, Sezione di Bologna, viale Berti Pichat 6/2, 40127, Bologna, Italy}
 
 \author{Alessandro Granelli}
 \email{alessandro.granelli@unibo.it}
\affiliation{Dipartimento di Fisica e Astronomia, Università di Bologna, via Irnerio 46, 40126, Bologna, Italy;}
 \affiliation{INFN, Sezione di Bologna, viale Berti Pichat 6/2, 40127, Bologna, Italy}
 
\author{Jacopo Nava}
\email{jacopo.nava2@unibo.it}
\affiliation{Dipartimento di Fisica e Astronomia, Università di Bologna, via Irnerio 46, 40126, Bologna, Italy;}
 \affiliation{INFN, Sezione di Bologna, viale Berti Pichat 6/2, 40127, Bologna, Italy}
 
 \author{Filippo Sala}
 \email{f.sala@unibo.it}\thanks{FS is on leave from LPTHE, CNRS \& Sorbonne Universit\'{e}, Paris, France.}
\affiliation{Dipartimento di Fisica e Astronomia, Università di Bologna, via Irnerio 46, 40126, Bologna, Italy;}
 \affiliation{INFN, Sezione di Bologna, viale Berti Pichat 6/2, 40127, Bologna, Italy}

 \begin{abstract}

\noindent Models of blazar jets, that explain observations of their photon spectra, typically predict too few neutrinos to be possibly seen by existing telescopes. In particular, they fall short in reproducing the first neutrino ever detected from a blazar, TXS 0506+056, by IceCube in 2017.
We predict larger neutrino fluxes by using the same jet models, extended to include deep inelastic scatterings between protons within the jets and sub-GeV dark matter (DM) around the central black holes of blazars.
In this way we succeed in explaining neutrino observations of TXS 0506+056, for DM parameters allowed by all laboratory, direct and indirect searches. Our proposal will be tested by DM searches, as well as by the observation of more neutrinos from blazars.
Our findings motivate to implement DM-nuclei interactions in jet models and to improve our knowledge of DM spikes around active galactic nuclei.

 \end{abstract}
\maketitle

\noindent \textit{\textbf{Introduction}}--- The existence of dark matter (DM) in our Universe is well established,
on sub-galactic to cosmological scales. Its constituents, origin and non-gravitational interactions remain today an
outstanding mystery and are the object of intense investigation~\cite{Cirelli:2024ssz}.
Active galactic nuclei (AGN) are the most powerful steady luminous sources in the Universe,
fueled by supermassive black holes (BHs) in their center \cite{Padovani:2017zpf}.
If they feature visible jets of relativistic particles and one of them is in close alignment with our line-of-sight (LOS)
then AGN are called blazars~\cite{Padovani:2017zpf, Urry:1995mg}. 
Their properties are still not well understood and the subject of active scrutiny, propelled by many observations of blazar photons and by the tentative detection of correlated photon and neutrino emission from the blazar TXS 0506+056~\cite{IceCube:2018dnn}.
Given that extremely high densities of DM are expected to accumulate into \virg{spikes} around black holes~\cite{Gondolo:1999ef}, and that the nature of both DM and AGN jets is still elusive,
it appears natural to inquire what such observations can teach us about DM via AGN and viceversa.

This direction has been pioneered in~\cite{Gorchtein:2010xa}, which studied the DM effects on photons from AGN.
It has recently been revived in~\cite{Wang:2021jic,Granelli:2022ysi}, which proved that the DM fluxes upscattered by blazars' jets could induce recoils on Earth, that lead to promising sensitivities on Standard Model (SM) interactions of sub-GeV DM.
The energy of DM constituents within the Milky Way halo is too low to induce nuclear recoils detectable with the leading \virg{direct detection} experiments, if their mass $\MDM \lesssim\,\text{few GeV}$.
The best sensitivities in that mass range then come either from new detection techniques of
halo DM~\cite{Essig:2022dfa}, or from searches with DM~\cite{Bringmann:2018cvk,PandaX:2023tfq} or
neutrino experiments~\cite{Ema:2020ulo,Super-Kamiokande:2022ncz} for the subpopulation of energetic DM that necessarily exist,
e.g.~DM upscattered by galactic cosmic rays~\cite{Bringmann:2018cvk,Ema:2018bih} or from atmospheric production~\cite{Alvey:2019zaa}.
While blazar upscattering admittedly suffers from larger astrophysical uncertainties than the latter two mechanisms,
it is well worth exploring because it could be the first one leading to DM detection,
and because it offers unique experimental opportunities thanks to its distinct temporal, spatial and energetic properties.

In this paper we explore, for the first time, the impact on blazar neutrinos of DM-SM interactions in blazar jets.
This direction is additionally encouraged by existing and planned observations of high-energy blazar neutrinos and by their current understanding.
Indeed, while existing models of blazar jets allow to explain their photon emission across a vast
range of energies, they typically predict too few neutrinos to be possibly seen by existing telescopes,
see e.g.~\cite{Cao:2023wug}.
In particular, different models of TXS 0506+056 jets all
predict~\cite{Gao:2018mnu,Cerruti:2018tmc,Cerruti:2018tmc_erratum,Keivani:2018rnh}
a neutrino flux roughly two orders of magnitude lower than IceCube's observations,
%too low to satisfactorily explain IceCube's observations,
strongly motivating our investigation of its possible DM origin.\\

%%%%%%%%%%%%%%%%%%%%%%%%%%%%%%%%%%%%%%%%%%%%%%%%%%

\noindent \textit{\textbf{Blazar jet physics and source selection}}--- Blazars exhibit a distinctive
double-peaked spectral energy distribution (SED) of photons \cite{Blandford:2018iot}
that can be attributed to the non-thermal radiative emission
from charged particles in the jets as they propagate
through magnetic fields and ambient radiation.
Typically, the emitting regions, in which the particles in the jets are confined, are considered as spherical \textit{blobs} 
\cite{Dermer:2009zz}
that move
with Lorentz 
factor $\Gamma_B$ 
along the jet axis, the latter being
inclined by an angle $\theta_{\LOS}$ with respect to the LOS. For the particles in the blob,
it is customary to consider homogeneous and isotropic power-law energy spectra.
Here we shall concentrate on protons $p$ and express their spectrum as
$ d\Gamma'_p/(d\gamma'_p d\Omega')=(\kappa_p/4\pi){\gamma_p'}^{-\alpha_p}$,
where $d\Gamma'_p$ is the infinitesimal rate of protons ejected in the blob
in the direction $d\Omega'$ with Lorentz factor $\gamma'_p$ in the range $[\gamma'_p, \gamma'_p + d\gamma'_p]$,
$\alpha_p$ is the power-law slope, and
$\kappa_p$ is a normalisation constant. 
Hereafter, primed (unprimed) quantities refer to the blob's (observer's) rest frame.

A compelling class of models for the jet emission is that of lepto-hadronic ones, 
where both electrons and protons are accelerated to extreme velocities and
the highest-energy blazar photons are generated
via a combination of leptonic and hadronic processes (see, e.g., \cite{Cerruti:2020lfj} for a review).
By allowing for ultra-relativistic protons, these models have the
intrinsic property to predict high-energy secondary neutrinos
from photo-meson and proton-proton interactions. 
Lepto-hadronic models attracted increasing attention after IceCube reported the first spatial association between neutrino events and a blazar,  TXS 0506+056, with a significance larger than 3$\sigma$ \cite{IceCube:2018dnn, IceCube:2018cha, Padovani:2018acg}.
Specifically, there have been two such associations: one related to the single neutrino event in 2017 (IC-170922A) in coincidence with a six-month 
multi-band flare of TXS 0506+056 \cite{IceCube:2018dnn}, and $\sim 13$ events prior to IC-170922A in 2014/2015 \cite{IceCube:2018cha}, which, however, were not accompanied by an enhanced electromagnetic activity of the aforementioned blazar \cite{Fermi-LAT:2019hte}. 
Following these events, various hybrid lepto-hadronic models for TXS 0506+056 have been tested against the observed SED and neutrino flux \cite{Gao:2018mnu, MAGIC:2018sak, Cerruti:2018tmc,Cerruti:2018tmc_erratum, Keivani:2018rnh, Liu:2018utd, Righi:2018xjr, Sahakyan:2018voh, Rodrigues:2018tku, Zhang:2019dob, Xue:2019txw, Oikonomou:2019djc, Petropoulou:2019zqp, Gasparyan:2021oad}.
While they manage to explain the photon's SED, they typically predict a too small neutrino flux to explain the observed events, motivating the possibility that they have a different origin (see e.g.~\cite{KhateeZathul:2024tgu}).
Other blazars have been correlated with astrophysical neutrinos, 
although with smaller significance compared to TXS 0506+056 (see e.g.~\cite{Kadler:2016ygj, Paliya:2020mqm, Rodrigues:2020fbu, Oikonomou:2021akf, Liao:2022csg, Sahakyan:2022nbz, Fermi-LAT:2019hte, Jiang:2024nwa, Ji:2024dgn, Ji:2024zbv}, also \cite{Giommi:2021bar, Boettcher:2022dci} for reviews). 
This motivates lepto-hadronic models as frameworks to describe the whole population of blazars.

In this work we use lepto-hadronic models of blazar jets as an input.
For TXS 0506+056 we use first the model of~\cite{Keivani:2018rnh} that yields the largest neutrino flux, and then the one of~\cite{Cerruti:2018tmc,Cerruti:2018tmc_erratum}. To investigate whether our findings are peculiar to TXS 0506+056 or more general, we also consider the subset of all the blazars modeled in
 \cite{Rodrigues:2023vbv}, whose neutrino flux at Earth is constrained by the IceCube stacking analysis in~\cite{Abbasi:2022uox}.
The study in \cite{Abbasi:2022uox} searched for astrophysical neutrinos from 137 blazars in the first \textit{Fermi}-LAT low-energy catalog (1FLE), using ten years of IceCube muon-neutrino data.
After identifying
the sources included in both analyses~\cite{Abbasi:2022uox, Rodrigues:2023vbv},
we select AP Librae as a representative one.

The jet parameters that are relevant to our analysis are: 
the minimal and the maximal boost factors of the protons in the blob frame $\gamma'_{\text{min}, p}$, $\gamma'_{\text{max}, p}$;
the spectral index $\alpha_p$; the Lorentz bulk factor $\Gamma_B$; 
the LOS angle $\theta_{\LOS}$; 
and the proton luminosity $L_p = \kappa_p m_p \Gamma_B^2 \int_{\gamma'_{\min,p}}^{\gamma'_{\max,p}}\,x^{1-\alpha_p}dx$ 
\cite{Granelli:2022ysi}, $m_p \simeq 0.938 \,\text{GeV}$ being the proton mass. The parameters for both TXS 0506+056 and AP Librae are summarised in Table \ref{tab:AGNparameters}, together with
their values for the redshift $z$ \cite{Paiano:2018qeq, Disney:1974abc, Stickel:1993abc, Jones:2009abs}, the luminosity distance $d_L$ (computed assuming standard cosmology \cite{ParticleDataGroup:2024cfk}),
the BH mass \cite{Padovani:2021kjr, Woo:2005abc} and the corresponding Schwarzschild radius $R_S$.
We impose an exponential suppression of the proton spectrum at energies larger than their maximal ones.\\

\setlength{\tabcolsep}{19pt}
\begin{table}
\centering
\begin{tabular}{@{}c||cc@{}}
    \hline
    \rowcolor[gray]{.95}
    \multicolumn{1}{@{}c@{}||}{\textbf{Parameter}}&
    \multicolumn{1}{@{}c@{}}{\textbf{TXS 0506+056}} &
     \multicolumn{1}{@{}c@{}}{\textbf{AP Librae}} \\
    \hline
    \hline
    \rule{0pt}{2.5ex}~~~~~~$z$ & 0.337 & 0.05\\
 ~~~~~~$d_L$ (Mpc) & 1774.92 & 223.7 \\
 ~~~~~~$M_{\BH}\,(M_\odot)$  & $3\times 10^8$ & $3\times 10^8$\\
    ~~~~~~$R_S$ (\text{pc}) & $3\times 10^{-5}$ & $3\times 10^{-5}$ \\
 ~~~~~~$\Gamma_B$ & $24.2$ & 4\\
 ~~~~~~${\theta_{\LOS}}\,(^{\circ})$ & 2.37 & 14.3\\
 ~~~~~~$\alpha_p$ & 2 & 1\\
 ~~~~~~$\gamma'_{\min, p}$ & 1 & 100\\
 ~~~~~~$\gamma'_{\max, p}$ & $1.6 \times 10^{7}$ & $1.3 \times 10^{7}$\\
 ~~~~~~$L_p$ (erg/s)  & $1.85\times 10^{50}$ & $3.18 \times 10^{47}$ \\
 ~~~~~~$\kappa_p\,(\text{s}^{-1}\text{sr}^{-1})$ & $1.27 \times 10^{49}$ & $1.01 \times 10^{42}$\\ 
\hline
\end{tabular}
\caption{The relevant parameters from lepto-hadronic fits for the blazars TXS 0506+056 \cite{Keivani:2018rnh} and  
AP~Librae~\cite{Rodrigues:2023vbv} 
used in our calculations. Also listed are the redshift $z$ \cite{Paiano:2018qeq, Disney:1974abc, Stickel:1993abc, Jones:2009abs}, the luminosity distance $d_L$ \cite{ParticleDataGroup:2024cfk} and
the BH mass \cite{Padovani:2021kjr, Woo:2005abc} in solar mass units ($M_\odot$),
and the Schwarzschild radius $R_S$.}
\label{tab:AGNparameters}
\end{table}

\begin{figure*}[t]
    \centering
    \includegraphics[width=0.49\textwidth]{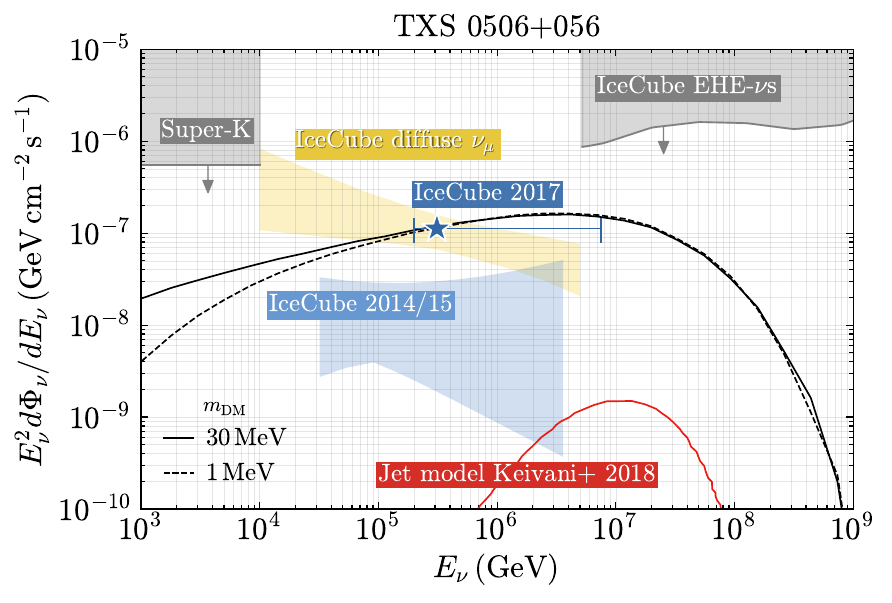}\,
    \includegraphics[width=0.49\textwidth]{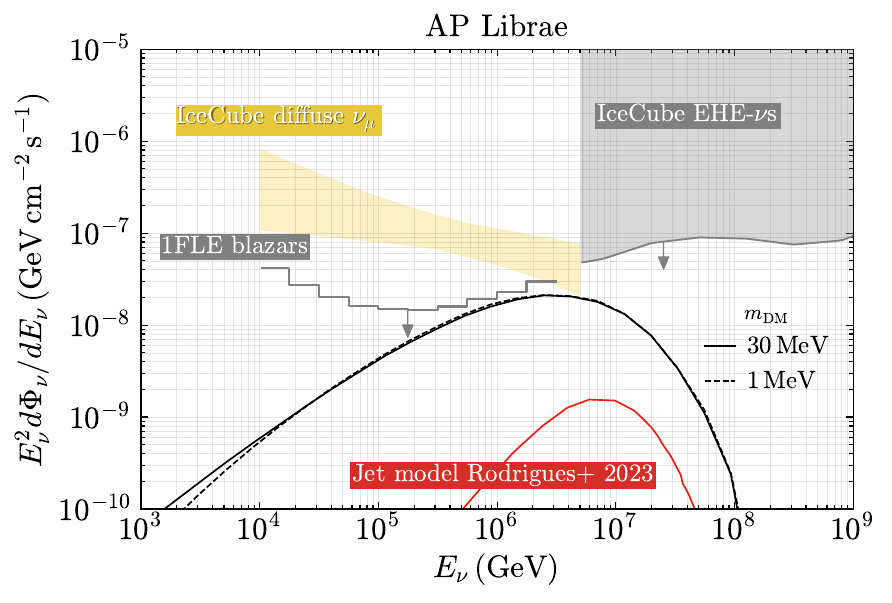}
    \caption{Our calculations of the differential neutrino flux,
    from DIS between protons in a blazar jet and
    DM around the BH fueling it, are displayed as \textbf{black (black dashed) lines}
    for $\MDM = 30~(1)$~MeV, for the blazars \textbf{TXS 0506+056} (\textbf{left}) and
    \textbf{AP Librae} (\textbf{right}). The neutrinos predicted by jet
    models~\cite{Keivani:2018rnh} (TXS 0506+056)
    and~\cite{Rodrigues:2023vbv} (AP Librae) are displayed as \textbf{red lines}. 
    IceCube detections of neutrinos from TXS 0506+056 are displayed
    as a \textbf{blue segment} (2017 event~\cite{IceCube:2018dnn}, $90\%$ C.L.~in energy, best-fit marked by a \textbf{star})
    and \textbf{blue band} ($95\%$ C.L.~$\nu_\mu+\overline{\nu}_\mu$
    during the 110 days flare 2014/15~\cite{IceCube:2018cha}, which we rescale to 6 months).
    In both plots we report as \textbf{yellow bands}, for comparison,
    the diffuse $\nu_\mu + \bar{\nu}_\mu$ detected by IceCube~\cite{Abbasi:2021qfz} at $95\%$ C.L.
    The \textbf{grey lines} with downward arrows represent the $90\%$ C.L.~upper limits from Super-K towards TXS 0506+056~\cite{Super-Kamiokande:2019utr} (left), % and XYX~\cite{} (right),
    IceCube 1FLE blazar searches Fig.~2 of~\cite{Abbasi:2022uox} (right), see also~\cite{Reimann:2019zby},
    and IceCube using nine-years null-observations of extremely-high-energy neutrinos~\cite{IceCube:2018fhm} (on the left rescaled to 6 months).}
    \label{fig:nu_fluxes}
\end{figure*}

%%%%%%%%%%%%%%%%%%%%%%%%%%%%%%%%%%%%%%%%%%%%%%%%%%
\noindent \textit{\textbf{Dark matter around blazars}}--- In the proximity of  supermassive BHs, like those fueling blazars,
DM accumulates into spikes. In \cite{Gondolo:1999ef}, Gondolo $\&$ Silk (GS) demonstrated that, if the BH evolves adiabatically in the center of a spherical DM halo with density $\rho^\text{halo}_{\DM}(r)=\mathcal{N} r^{-\gamma}$, the halo grows in its center a spike with density
 $\rho^\text{spike}_{\DM}(r) = \mathcal{N} R_\text{sp}^{-\gamma} (R_\text{sp}/r)^{\alpha_{\GS}}$, with $\alpha_{\GS}=(9-2\gamma)/(4-\gamma)$,
$\mathcal{N}$ a normalisation constant, $r$ the
radial distance from the central BH, $R_\text{sp} \simeq \epsilon(\gamma) (M_{\BH}/\mathcal{N})^{1/(3-\gamma)}$ the extension of the spike 
and $\epsilon(\gamma)\approx 0.1$ for $0.5\leq \gamma \leq 1.5$ \cite{Merritt:2003qc}.
For the blazars under consideration, we model the total DM profile $\rho_{\DM}(r)$ as $\rho_{\DM}(r \geq R_\text{sp}) = \rho_{\DM}^\text{halo}(r)$ with $\gamma = 1$, as for a Navarro-Frenk-White (NFW) distribution \cite{Navarro:1995iw, Navarro:1996gj},
whereas $\rho_{\DM}(2R_S \leq r < R_\text{sp}) = g(r) \rho_{\DM}^\text{spike}(r)$ with $\alpha_{\GS} = 7/3$, and $g(r)=(1-2R_S/r)^{3/2}$ accounting for 
the inevitable capture of DM onto the BH~\cite{Gondolo:1999ef}, relativistic effects included \cite{Sadeghian:2013laa}.

The scarcity of information on the DM distribution around blazars
renders the normalisation of the DM profile somewhat arbitrary.
In our analysis, we fix $R_\text{sp} = R_\star$, where $R_\star\approx 10^6R_S$ is the typical radius of influence of a BH on stars \cite{Kormendy:2013dxa}.
This normalisation results in $\mathcal{N} \simeq 3\times 10^{-6} M_\odot/R_S^2$ for the blazars under consideration.
It ensures that within $R_\star$ the DM amounts to $\mathcal{O}(10\%)\,M_{\BH}$,
without interfering with BH mass estimates \cite{Labita:2006jg, Pei:2021qay}.
Note that our normalisation results in less DM around the BH than in other DM-AGN studies~\cite{Gorchtein:2010xa, Wang:2021jic, Granelli:2022ysi, Cline:2022qld, Ferrer:2022kei, Bhowmick:2022zkj, Cline:2023tkp, Herrera:2023nww} and is thus more conservative.

Many effects can influence the formation and evolution of a DM spike, like galaxy mergers~\cite{Ullio:2001fb, Merritt:2002vj},
the gravitational interaction of stars close to the BH~\cite{Gnedin:2003rj, Bertone:2005hw}, and DM annihilations over the BH lifetime $t_{\BH}$~\cite{Gondolo:1999ef}.
The latter implies that the DM density softens its slope to $\leq 0.5$ for $r<r_\text{ann}$~\cite{Shapiro:2016ypb}, where $r_\text{ann}$ is defined by $ \rho_{\DM}(r_\text{ann}) = \MDM/(\langle\sigma_\text{ann} v_\text{rel}\rangle t_{\BH})$,
and $\langle\sigma_\text{ann} v_\text{rel}\rangle$ is the DM averaged annihilation cross section times relative velocity.

For our calculations, all the information
on the DM distribution ultimately condenses into the following LOS integral \cite{Wang:2021jic, Granelli:2022ysi} (see also \cite{Profumo:2013jeb}):
\begin{equation}
    \Sigma_{\DM}^\text{spike} \equiv \int_{r_{\min}}^{R_\text{sp}} \rho_{\DM}(r') dr',
\end{equation}
with $r_{\min}$ being the minimal radial extension of the blazar jet. 
Admittedly, large uncertainties reside in $r_{\min}$,
with this having a substantial impact on $\Sigma_{\DM}^\text{spike}$.
We find it then useful to use the GS spike and consider different benchmark cases (BMCs) for $r_{\min}$,
as a way to effectively account also for astrophysical or DM softenings of the spike.
In particular, we choose
$r_{\min} = 10^2 R_S$ (BMCI) and $r_{\min} = 10^4 R_S$ (BMCII), 
corresponding to distances at which blazar studies expect their jets to be already well-accelerated \cite{Rodrigues:2023vbv}.
Within the adopted normalisation, we find $\Sigma_{\DM}^\text{spike} \simeq 6.9\times 10^{28} \,(1.5\times 10^{26})\,\text{GeV}\,\text{cm}^{-2}$ for BMC I (BMC II). 
We check that $r_\text{ann} < r_\text{min}$ allows for $\left\langle \sigma_\text{ann} v_\text{rel}\right\rangle\lesssim 1.4\times 10^{-25}\,(3.1\times 10^{-30})\,\text{cm}^3\,\text{s}^{-1}(m_{\DM}/\text{GeV})$, for $t_{\BH} = 10^{9}\,\text{yr}$.
We provide more details on $\Sigma_{\DM}^\text{spike}$ in Appendix~\ref{app:DMprofile}.\\

%%%%%%%%%%%%%%%%%%%%%%%%%%%%%%%%%%%%%%%%%%%%%%%%%%

\noindent \textit{\textbf{Neutrino flux from dark matter-proton scatterings}}--- If DM interacts with hadrons,
protons in the relativistic jets of a blazar can collide with the DM particles along their way. These collisions have a centre-of-mass energy
$\sqrt{s}
\sim \sqrt{2m_{\DM} m_p \gamma_p}$, which can be much larger than a GeV, depending on $m_{\DM}$.
At these energies,
the DM-proton scattering is dominated by the inelastic contribution. 
When protons scatter inelastically with DM, they disintegrate and
generate hadronic showers, with subsequent production of 
several final-state neutrinos from meson and secondary lepton decays.

We estimate the resulting neutrino flux (per-flavour) at Earth as
\begin{equation}
\frac{d\Phi_\nu}{dE_\nu} \simeq 
\frac{1}{3}\frac{\Sigma^\text{spike}_{\DM}}{\MDM d^2_L}
\int^{\gamma_p^{\max}}_{\gamma_p^{\min}(E_\nu)} \frac{d\Gamma_p}{d\gamma_p d\Omega}
\biggr\rvert_{\theta_{\LOS}}
\!\!\left<\frac{dN_\nu}{dE_\nu}\right>
\sigma^{\DIS}_{\DM-p}
d\gamma_p,
\end{equation}
where $\gamma_p^{\min}(E_\nu)$ is the minimum proton boost factor 
necessary to produce a neutrino of energy
$E_\nu$, dependent also on $\MDM$;
$d\Gamma_p/(d\gamma_p d\Omega)$ is the differential proton flux in the observer's frame (see, e.g., \cite{Wang:2021jic, Granelli:2022ysi});
$\sigma^{\DIS}_{\DM-p}$ 
is the 
integrated DM-proton deep inelastic scattering (DIS) cross section;
the $1/3$ factor accounts for neutrino oscillations over extragalactic distances,
which homogenise the neutrino flux among the three flavors; 
$\left<dN_\nu/dE_\nu\right>$ 
is
the number of neutrinos produced per $E_\nu$
at a given $s$, averaged over all the possible scatterings with the quarks, 
each weighted by its respective differential cross section. 
To compute
$\left<dN_\nu/dE_\nu\right>$, 
we have implemented 
the considered model of
the DM-proton interaction (see further) 
in
\textsc{FeynRules} \cite{Alloul:2013bka},
imported it in \textsc{Madgraph5}~\cite{Alwall:2014hca}
and simulated the collisions. 
We then generated the showering,
hadronization and neutrino emission using \textsc{Pythia8}~\cite{Bierlich:2022pfr}. 
Due to the extreme energies involved, 
the outgoing neutrinos in the observer's frame 
are predominantly collinear with the incoming protons.
Hence, 
we consider only a small cone around the LOS 
within which 
$d\Gamma_p/(d\gamma_p d\Omega) \simeq [d\Gamma_p/(d\gamma_p d\Omega)]_{\theta_{\LOS}}$.

Our predicted neutrino fluxes are shown in Fig.~\ref{fig:nu_fluxes} for TXS 0506+056 and AP Librae.
They have both a larger amplitude and a different shape than the neutrino fluxes predicted, independently of DM,  by the jet models
of the same blazars, also shown in Fig.~\ref{fig:nu_fluxes}.
We display in the same figure various neutrino observations and limits, including the diffuse $\nu_\mu + \bar{\nu}_\mu$ detected by IceCube~\cite{Abbasi:2021qfz}\footnote{Note that Fig.~\ref{fig:nu_fluxes} does not imply that TXS 0506+056 can power the diffuse flux, because the latter does not come only from a specific direction, and because TXS 0506+056 was in flare for 6 months, while the band of diffuse IceCube neutrinos comes from observations over a period of 9.5 years.}.
Our proposal of DM-proton scatterings can explain the neutrinos observed by IceCube from TXS 0506+056.
At a given value of $\MDM$, the choice of a specific DM model and interaction strength just changes the overall flux normalisation, not its shape.
The shape of the flux depends on $\MDM$ only at small $E_\nu$, where protons are not energetic enough to break on DM.
More details and checks about our computation of the neutrino fluxes are given in Appendix~\ref{app:flux}.

We prove next that the DM parameters corresponding to the normalisations displayed in Fig.~\ref{fig:nu_fluxes}
are allowed by all existing DM searches. For TXS 0506+056, only the results for the fit of \cite{Keivani:2018rnh}
are presented in Fig.~\ref{fig:nu_fluxes},
while those for the fit of \cite{Cerruti:2018tmc,Cerruti:2018tmc_erratum} are described in Appendix~\ref{app:TXS_Cerruti}.\\

\begin{figure}[t]
    \centering
    \includegraphics[width=0.5\textwidth]{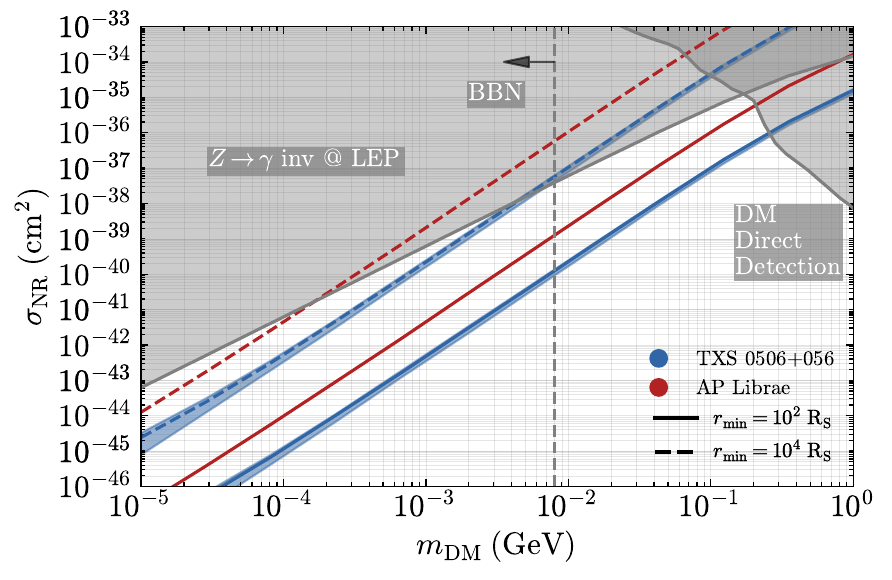}
    \caption{DM parameter space of Eqs~(\ref{eq:vector}), (\ref{eq:sigma}), for $m_V = 5$~GeV.
    The 2017 neutrino detected from TXS 0506+056~\cite{IceCube:2018dnn}
    is explained along the blue lines ($E_\nu$ best-fit)
    and shaded regions ($90\%$ C.L.~in $E_\nu$), and the upper limits from blazars~\cite{Abbasi:2022uox} (including AP Librae) are respected below the red lines. Continuous (dashed) lines correspond to BMCI (BMCII) for the DM spike.
    The grey shaded areas are excluded by direct detection of halo DM at SuperCDMS~\cite{SuperCDMS:2023sql}, SENSEI~\cite{SENSEI:2023zdf}, CRESST-III \cite{CRESST:2019jnq}, DarkSide-50~\cite{DarkSide:2018bpj}, XENONnT~\cite{XENON:2023cxc}, and by LEP searches for $Z\to \gamma \text{ invisible}$~\cite{L3:1997exg,DELPHI:1996qcc} in the UV completion of~\cite{Dror:2017ehi}. On the left-hand side of the vertical grey line, BBN excludes the model unless either a DM coupling to neutrinos is added~\cite{Escudero:2018mvt,Sabti:2019mhn} or DM is frozen-in below the QCD scale~\cite{Berlin:2018ztp}.}
    \label{fig:DM}
\end{figure}
\noindent \textit{\textbf{Dark matter-proton interaction}}---
We consider DM as a SM singlet Dirac fermion $\chi$, coupled to the first-generation quarks $q=u,d$ via a vector mediator $V$ with mass $m_V$.
The low-energy Lagrangian reads
\begin{equation}
\label{eq:vector}
 \mathcal{L}_{V} = g_{\chi V} \bar{\chi}\gamma^\mu \chi V_\mu + g_{q V} \bar{q} \gamma^\mu q V_\mu \,,
\end{equation}
where $g_{\chi V}$ and $g_{q V}$ are dimensionless couplings. In presenting our results,
we trade them for the non-relativistic spin-independent
DM-proton cross section
\begin{equation}
\label{eq:sigma}
\sigma_\text{NR}=\frac{g_{\chi V}^2 g_{pV}^2}{\pi}\frac{\mu_{\chi p}^2}{m_V^4}\,, \hspace{1cm} g_{pV}=2 g_{uV} + g_{dV} \,,
\end{equation}
where $\mu_{\chi p}=m_{\DM} m_p/(m_{\DM} + m_p)$ and we assume $g_{u V}=g_{d V}=g_V$ and $m_V = 5$~GeV for definiteness.
Larger values of $m_V$ exceed the typical transferred momentum, suppressing our signal;
$m_V < 5$~GeV would enlarge the parameter space where our proposal works, because of the additional contribution of QCD resonances, which we leave, however, to future work.

In Fig.~\ref{fig:DM}, lines indicate the values of DM parameters that give rise to the neutrino fluxes of Fig.~\ref{fig:nu_fluxes} from DM-proton scatterings within the blazar jets of TXS 0506+056 and AP Librae. We also display the strongest existing limits that, to our knowledge,
come from direct detection (DD)~\cite{SuperCDMS:2023sql,SENSEI:2023zdf,CRESST:2019jnq,DarkSide:2018bpj,XENON:2023cxc}, LEP searches for $Z\to \gamma V$ with $V$ decaying invisibly~\cite{L3:1997exg,DELPHI:1996qcc, Dror:2017ehi} and Big Bang Nucleosynthesis (BBN)~\cite{Escudero:2018mvt,Sabti:2019mhn}.
While BBN and DD limits are model-independent, laboratory ones depend on the UV completion.
For concreteness we show the laboratory limits associated to a UV completion with gauged baryon number, following~\cite{Dror:2017ehi}.
Different choices of $m_V$ do not affect BBN nor DD, while LEP limits get more stringent as $m_V$ decreases. We checked that we are able to explain the neutrino from TXS 0506+056, for BMCI and compatibly with laboratory limits, for $m_V$ down to 400~MeV, where kaon decays open up.
We discuss how our findings differ for a DM model with scalar mediator in Appendix~\ref{app:ScalarMediator} (see also \cite{DeMarchi:2025xag} 
for even more cases).\\

%%%%%%%%%%%%%%%%%%%%%%%%%%%%%%%%%%%%%%%%%%%%%%%%%%

\noindent \textit{\textbf{Summary and discussion}}---
TXS 0506+056 is the first blazar from which a PeV neutrino has been tentatively detected, by IceCube in 2017~\cite{IceCube:2018cha}.
Models of TXS 0506+056 jets, that explain the photons also observed from it over a large energy range,
fall short in reproducing IceCube’s observation by roughly two orders of magnitude.
In this paper, by using the same jet models, we proved that the high-energy neutrinos detected from TXS 0506+056
could originate from deep inelastic scatterings between protons within its jet and sub-GeV dark matter around its central black hole, see Fig.~\ref{fig:nu_fluxes}.
We checked that our finding is robust, in the sense that it holds for different jet models that fit TXS 0506+056 photon emission.
We further proved that TXS 0506+056 is not a peculiar blazar in this sense, but that also for other blazars dark matter-proton scatterings can induce neutrino fluxes larger than those coming from their jets.

Importantly, we checked that the sub-GeV dark matter parameters leading to the above conclusions are allowed by all existing laboratory,
direct and indirect searches, and that they could soon be discovered there,
see Fig.~\ref{fig:DM}. Still on the dark matter side,
our work proves that dark matter interactions with protons in the
jets of blazars could be first observed via the deviations they induce
in the neutrino events from blazars, rather than by looking for the dark matter
that is itself upscattered by the blazar as proposed in~\cite{Wang:2021jic}.
In a forthcoming publication \cite{DeMarchi:2025uoo} we prove that this conclusion still holds when one
improves over studies like~\cite{Wang:2021jic} by considering the same blazars and dark matter models considered here,
and larger detectors than XENONnT for the detection of blazar-upscattered dark matter.

Our findings motivate several avenues of investigation.
First, blazar neutrinos from dark matter-proton scatterings have not only a potentially substantial amplitude,
but also a distinct energy shape.
The association of more neutrinos to blazars will then offer an immediate observational ground to test our proposal versus other explanations of those neutrinos.
In this sense, it will be intriguing to study the hints of other neutrinos from blazars~\cite{Kadler:2016ygj, Paliya:2020mqm, Rodrigues:2020fbu, Oikonomou:2021akf, Liao:2022csg, Sahakyan:2022nbz, Fermi-LAT:2019hte, Jiang:2024nwa, Ji:2024dgn, Ji:2024zbv},
as well as the $E_\nu \gtrsim \mathcal{O}(10)$~PeV neutrino detected by KM3NeT~\cite{CoelhoJ_presentation}.
On the modelling side, more accurate predictions need the implementation of deep inelastic scatterings at momentum transfer $\sim\text{GeV}$,
like the effect of resonances, and the inclusion of dark matter-proton scatterings in the fits of blazar jets. Such inclusion would also enable progress in several new directions, for example to check that the good astrophysical understanding of blazar photons is preserved. It would also make it possible to address one of the major shortcomings of lepto-hadronic jet models, i.e. their too large proton luminosities, thanks to the contribution of DM-proton scatterings to the blazar neutrino flux.
Finally, our conclusions call out loud for a better understanding of dark matter clustering around active galactic nuclei.

Our results suggest the intriguing possibility that the first neutrino
detected from a blazar could %also
be the first sign of a non-gravitational
dark matter interaction. 
Progress on the observational and phenomenological sides will tell.\\

%%%%%%%%%%%%%%%%%%%%%%%%%%%%%%%%%%%%%%%%%%%%%%%%%%
\acknowledgments

\section*{Acknowledgements}
We thank Michael Campana, Yohei Ema and Jin-Wei Wang for discussions.
A.G.~is grateful to the Department of Physics at UC San Diego for the hospitality offered during the final stages of this project.
J.N.~acknowledges hospitality from the Fermilab Theoretical Physics Department during the early stages of this work.
 This work was supported in part by the European Union's Horizon research and innovation programme under the Marie Skłodowska-Curie grant agreements No.~860881-HIDDeN and No.~101086085-ASYMMETRY, by COST (European Cooperation in Science and Technology) via the COST Action COSMIC WISPers CA21106, and by the Italian INFN program on Theoretical Astroparticle Physics.

%%%%%%%%%%%%%%%%%%%%%%%%%%%%%%

\appendix
\section{More details on the dark matter column density}
\label{app:DMprofile}

The column density $\Sigma_{\DM}^\text{spike}$ for a GS spike in the absence of DM annihilations
can be computed analytically as
\begin{equation}
    \Sigma_{\DM}^\text{spike} \simeq 
        \dfrac{\epsilon(\gamma)^{3-\gamma}}{\alpha_{\GS}-1}\frac{M_{\BH}}{r_{\min}^2}\left(\dfrac{r_{\min}}{R_\text{sp}}\right)^{3-\alpha_{\GS}}
\end{equation}
where we have used the relation $\mathcal{N} = M_{\BH} [\epsilon(\gamma)/R_\text{sp}]^{3-\gamma}$ \cite{Gondolo:1999ef}.

In the presence of DM annihilations, the DM spike gets flattened to a central core.
The radius at which the flattening takes place, $r_\text{ann}$,
can be obtained from the relation $\rho_{\DM}(r_\text{ann}) = \rho_\text{core} = m_{\DM} / (\left\langle\sigma_\text{ann}v_\text{rel}\right\rangle t_{\BH})$,
which gives
    \begin{equation}
r_\text{ann} = R_\text{sp}\begin{cases}
    \left[\dfrac{M_{\BH}\epsilon(\gamma)^{3-\gamma}}{R_\text{sp}^3\rho_\text{core}}\right]^{1/\alpha_{\GS}}
     & \text{if } r_\text{ann} \leq R_\text{sp},\\
     \left[\dfrac{M_{\BH}\epsilon(\gamma)^{3-\gamma}}{R_\text{sp}^3\rho_\text{core}}\right]^{1/\gamma}
     & \text{if } r_\text{ann} > R_\text{sp},\\
    \end{cases}    
\end{equation}
The contribution to the column density from the core due to annihilations reads $\Sigma_{\DM}^\text{core} \simeq \rho_\text{core}r_\text{ann}$.

Requiring that $r_\text{ann} = r_{\min} < R_\text{sp}$ results in the following condition on the annihilation cross section:
\begin{equation}
    \left\langle \sigma_\text{ann}v_\text{rel}\right\rangle = \frac{R_\text{sp}^3}{t_{\BH}}\frac{m_{\DM}}{M_{\BH}}\frac{1}{\epsilon(\gamma)^{3-\gamma}}\left(\frac{r_{\min}}{R_\text{sp}}\right)^{\alpha_{\GS}} 
\end{equation}
Fixing $t_{\BH} = 10^{9}\,\text{yr}$, $R_\text{sp} = 10^6 R_S$, $\epsilon= 0.1$, $M_{\BH}=3\times 10^8 M_\odot$,
we get $ \langle \sigma_\text{ann} v_\text{rel}\rangle \simeq 1.4\times 10^{-25}\text{cm}^3\,\text{s}^{-1} (m_{\DM}/\text{GeV})$
for $r_{\min} = 10^4R_S$. Any $\langle \sigma_\text{ann} v_\text{rel}\rangle$ below this threshold would lead to a DM column
density between the values adopted for BMC I and II. For $r_{\min}$ and the other
values as above we get $ \langle \sigma_\text{ann} v_\text{rel}\rangle \simeq 3.1\times 10^{-30}\text{cm}^3\,\text{s}^{-1} (m_{\DM}/\text{GeV})$.

\section{Computation of the neutrino flux}
\label{app:flux}
We consider a proton $p$ with mass $m_p$ scattering inelastically off a Dirac fermion $\chi$, constituting DM, with mass $m_\DM$.
If the momentum transfer $Q^2$ is large enough, the DM particles interact directly
with the quarks that constitute the proton via DIS. 
We find the following expression for the vector-mediated DM-$p$ DIS cross section (see Eq.~\eqref{eq:vector} for the Lagrangian of the considered interaction):
\begin{equation}
    \begin{split}\label{eq:DISvector}
\frac{d\sigma^{\DIS}_{\DM-p}}{dxdy} = \frac{1}{4\pi}\frac{(g_{\chi V})^2 (Q^2)^3 }{[(Q^2)^2-4m_p^2m_{\DM}^2 x^2 y^2] (Q^2+m_{V}^2)^2} \times \\
\Bigg[y F_1(x, Q^2) + \frac{1}{xy}\left(1-y-\frac{m_p^2 x^2 y^2}{Q^2}\right)F_2(x, Q^2)\Bigg]
    \end{split}
\end{equation}
where we have defined the Lorentz invariant quantities relevant to the DIS as
$y \equiv p_p\cdot q/(p_p\cdot p_\DM)$ 
and $x \equiv Q^2/(2 p_p\cdot q)= Q^2/[(s-m_p^2-m_{\DM}^2)\, y]$,
with $p_\DM$ ($k_\DM$) and $p_q$ the momenta of the initial (final) DM and initial quark, respectively,
$p_p$ that of the incoming proton, $s = (p_p + p_\DM)^2$
the squared centre-of-mass energy, $q=p_\DM - k_\DM$ the 4-momentum transfer and $Q^2 = - q^2$.
The $F$-functions are given by
\begin{eqnarray}
    F_1(x, Q^2) &=&  \frac{1}{2}\sum_{a=q,\bar{q}}(g_{a V})^2\,f_{a}(x, Q^2),\\
    F_2(x, Q^2) &=& 2xF_1(x, Q^2).
\end{eqnarray}
with $f_{q(\bar{q})}^{N}$ being
the parton distribution functions (PDFs) for the (anti)quarks 
$q\,(\bar{q})$. According to the DM-$p$ interaction model discussed in the main text,
we consider the contribution only of the up and down quarks.
Note that the example of UV completion that we discussed,
where the baryon number is gauged and extra fermions are introduced to
cancel anomalies~\cite{Dror:2017ehi}, predicts a coupling to heavier
quarks of the same size of the one to up and down quarks.
Including it would lead to a signal larger by a few tens of percent, strengthening our conclusions.

\begin{figure}[t]
    \centering
    \includegraphics[width=0.5\textwidth]{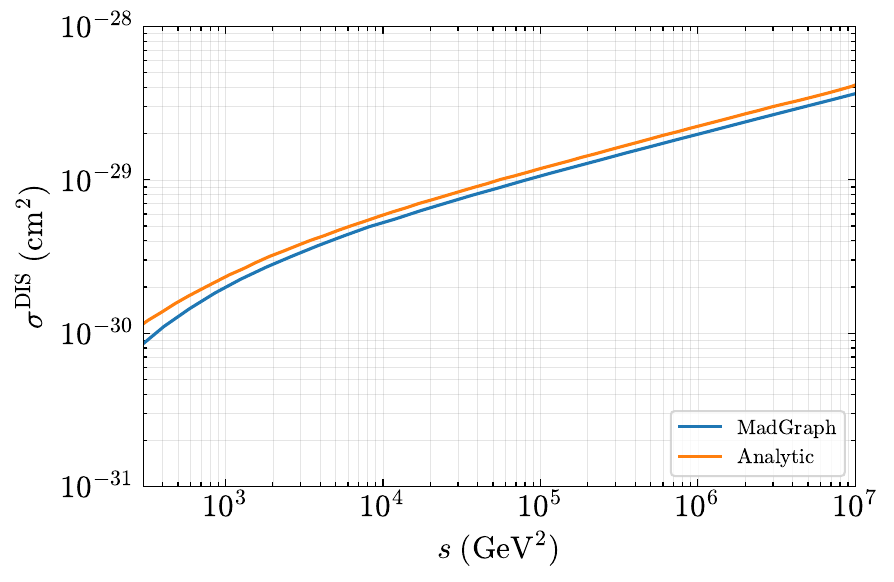}
    \caption{Comparison between our analytical computation of the cross section and the numerical evaluation performed by \textsc{Madgraph5}. Both are evaluated setting the couplings $g_\chi = g_q = 1$ and for $m_V = 5$ GeV.}
    \label{fig:sigma_analytical_MG_comparison}
\end{figure}
We have implemented our model of Eq.~(\ref{eq:vector})
in \textsc{FeynRules} and imported it into \textsc{Madgraph5}
to simulate the parton level scattering, using the PDF set
\virg{CT10} \cite{Lai:2010vv}. We
generate events for the process $\chi p \rightarrow \chi + \mathrm{jet}$,
in the $\chi p$ centre-of-mass frame. Because of the very large $p_p$
and $Q^2 \sim~\text{few GeV}^2$, and since $x$ can reach values of the order of
$Q^2/(\MDM p_p)$, it has been important to select PDFs extending to $x$
as small as $10^{-8}$ for TXS 0506+056 ($10^{-7}$ for AP Librae).
We have set all of \textsc{Madgraph5}'s kinematic cuts to zero,
as we are interested in all events that result in neutrinos,
regardless of their pseudorapidity or transverse momentum.
To avoid any divergent behaviour, \textsc{Madgraph5} imposes a
minimum momentum transfer $Q^2_{\mathrm{min}}\approx 4 \;\mathrm{GeV}^2$
below which the cross section is set to zero.
As a check of our results, we show in
Fig.~\ref{fig:sigma_analytical_MG_comparison} a
comparison between our analytical cross section
Eq.~\eqref{eq:DISvector}, using the same PDF set \virg{CT10}, 
with the one computed by \textsc{Madgraph5} at the
relevant $s\gtrsim \mathcal{O}(10^2)\,\text{GeV}^2$.
The agreement is at the level of roughly $20\%$, which is more than satisfactory for our purposes.
We believe the 
discrepancy between the two estimates
can be attributed to cuts that \textsc{Madgraph5} implements
internally and are not captured by the naive integration performed on the analytic cross section.  
We anyway make use of \textsc{Madgraph5}'s estimate of the cross section,
as it leads to a negligible underestimation of the flux.
Our calculation of the signal is additionally conservative
because we generate events at leading order, and thus neglect
all contributions from a gluon initial state, which we expect
to be non-negligible given the large gluon PDF at small $x$.
We recall, from the main text,
that we have not included the contribution to the scatterings from resonances,
which would be relevant at $Q^2 \sim$~GeV$^2$ and further increase our signal.

\begin{figure}[t!]
    \centering
        \includegraphics[width=0.5\textwidth]{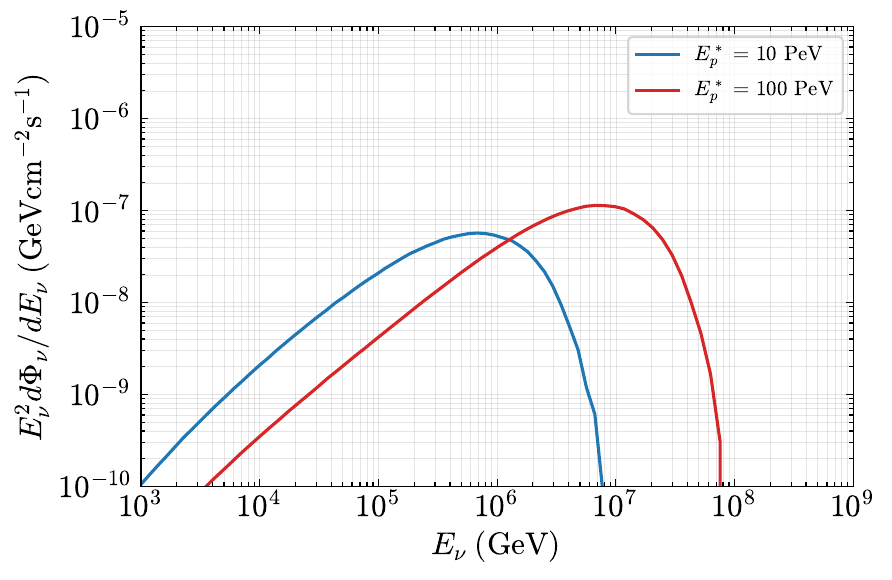}
        \caption{Neutrino flux in the case of a monochromatic jet, for two different values of the proton energy. The details of the normalisation are discussed in the text.}.
        \label{fig:monochromatic_jet}
    \end{figure}

We then pass the parton level scattering events generated by \textsc{Madgraph5} to \textsc{Pythia8},
to simulate the hadronization and subsequent decay process.
We select the final-state neutrinos and boost them from the $\chi p$ centre-of-mass
frame to the observer's frame, along the direction of motion of the proton, which we denote as $\hat{z}$.
The two are related via a Lorentz transformation with boost parameter
$\gamma_\mathrm{COM} = (m_\chi + E_p)/\sqrt{s}$.
The energy in the observer's frame, the $\hat{z}$ component of the neutrino's
momentum and its angle with respect to the $\hat{z}$ axis transform respectively as
\begin{equation}
\begin{aligned}
 E_\nu &= \gamma_\mathrm{COM}(E_\nu' + \beta p_{\nu, z}'),\\   p_{\nu, z} & = \gamma_\mathrm{COM}(p_{\nu, z}'+\beta E_{\nu}'), \\ 
 \cos\theta &= \frac{p_{\nu, z}}{\sqrt{{p'_{\perp}}^2+p_{\nu, z}^2   }},
\end{aligned}
\end{equation}
where the (un)primed quantities refer to the centre-of-mass (observer's)
frame, $p'_{\perp}$ is the component of the neutrino's 3-momentum
in the plane perpendicular to $\hat{z}$, and $\beta = \sqrt{1-1/\gamma_\mathrm{COM}^2}$.
We note that, for the purposes of our computation,
there is no need to extract the $Q^2$ and $x$ dependence
of the final-state neutrino distribution
$dN_\nu/dE_\nu$, as we only care about the final result of the convolution with the differential cross section,
integrated over $Q^2$ and $x$,
which is the final output from \textsc{Pythia8}.
That is, we are only interested in
\begin{equation}
\left< \frac{dN_\nu}{dE_\nu}\right>= \int_{x_\mathrm{min}}^{1}dx \; \int_{Q^2_{\mathrm{min}}}^{xs} dQ^2 \; \frac{1}{\sigma^{\DIS}_{\DM-p}}\frac{d^2\sigma^\mathrm{DIS}_{\DM-p}}{dQ^2 dx} \frac{dN_\nu}{dE_\nu},
\end{equation}
where $x_\mathrm{min} = Q^2_{\mathrm{min}}/s$ is the minimum fraction of proton momentum that the initial quark 
should carry to transfer a squared momentum $Q^2_\mathrm{min}$.% in the first place.

We also expand here on our handling of the angular dependence of the outgoing neutrinos.
As mentioned in the main text, after boosting from the centre-of-mass frame to the observer’s frame,
the angular distribution is extremely peaked in the direction collinear with the initial proton’s momentum.
This is due to the very high boost factor involved and means that the only neutrinos that will reach the Earth
come from protons that are already closely aligned with our LOS.
For this reason, we restrict our count of emitted neutrinos to a
narrow cone around the LOS, defined by an opening angle $\theta$ such that $1 - 10^{-5} \leq \cos\theta \leq 1$,
or equivalently $\theta < 0.0045\,\text{rad}$.
Within this small cone, the proton
spectrum is practically constant and it is safe to evaluate it in the direction of the LOS.

Finally, to allow for better reproducibility, we show in Fig.~\ref{fig:monochromatic_jet}
the neutrino flux in the case of a monochromatic blazar jet.
The proton fluxes in the figure are computed as
$d\Gamma_p/(d\gamma_p d\Omega) = \delta(E_p-E_p^*) m_p L_p/E_p^*$,
with proton luminosity $L_p = 1.85 \times 10^{50} \;\mathrm{erg}\mathrm{s}^{-1}$
and $E_p^* = 10, 100 \;\mathrm{PeV}$.
We also fix $g_\chi g_\mathrm{q} = 10^{-3}$ and
$\Sigma_\mathrm{DM} = 6.9\times 10^{28} \;\mathrm{GeV}\mathrm{cm}^{-2}$.

\noindent\section{Neutrino flux from TXS 0506+056 in alternative lepto-hadronic model}\label{app:TXS_Cerruti}
\begin{figure}[t!]
    \includegraphics[width=0.5\textwidth]{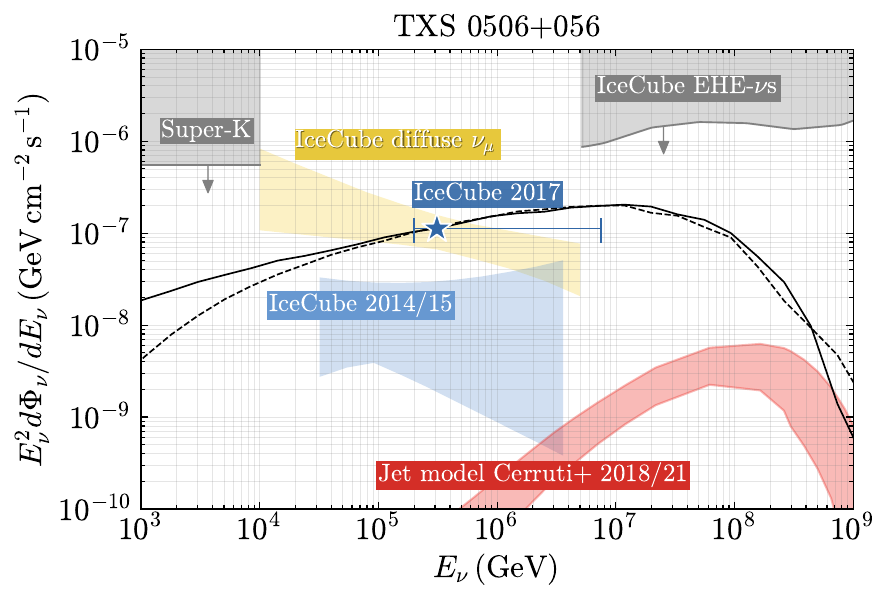}
    \includegraphics[width=0.5\textwidth]{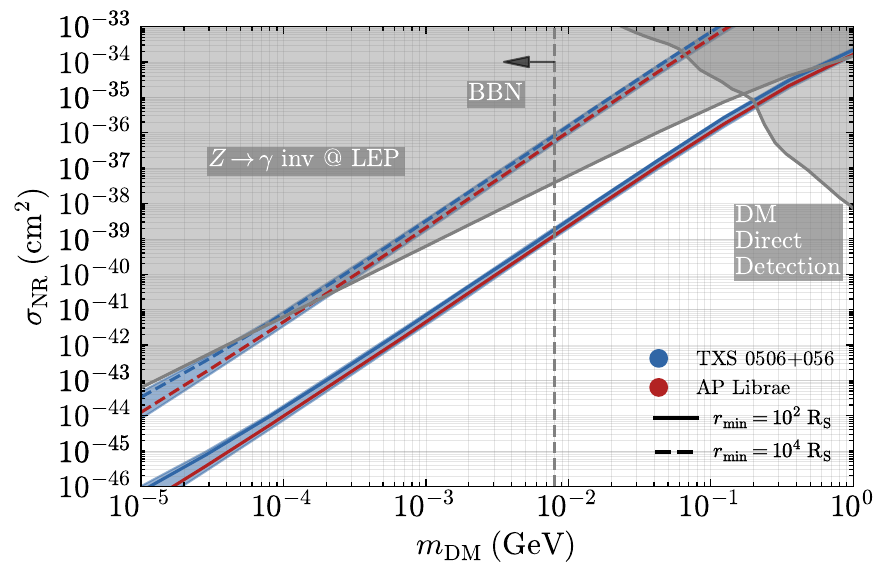}
    \caption{The upper (lower) panel is the same as in Fig.~\ref{fig:nu_fluxes} left panel (Fig.~\ref{fig:DM}), but for 
    the lepto-hadronic model of TXS 0506+056 presented in \cite{Cerruti:2018tmc,Cerruti:2018tmc_erratum} for the 2017 flare.}\label{fig:TXS_Cerruti}
    \end{figure}
We show in Fig.~\ref{fig:TXS_Cerruti} the neutrino flux resulting from
DM-proton DIS within the lepto-hadronic model presented in \cite{Cerruti:2018tmc,Cerruti:2018tmc_erratum}.
The best-fit parameters relevant to determine the TXS 0506+056 proton jet spectrum are \cite{Cerruti:2018tmc,Cerruti:2018tmc_erratum}:
$\Gamma_B = 20$, $\theta_{\LOS} = 0$, $\alpha_p=2$, $\gamma_{\min, p}' = 1$,
$\gamma_{\max, p}' = 5.5^\star\times 10^7$, $L_p = 2.55^\star\times 10^{48}\,\text{erg}\,\text{s}^{-1}$,
$\kappa_p = 2.39\times 10^{47}$. Starred quantities are computed as mean values of the ranges reported in 
\cite{Cerruti:2018tmc_erratum}. The results are analogous to the ones
discussed in the main text in the context of the lepto-hadronic model of \cite{Keivani:2018rnh}. This
proves that the conclusions of our work are robust with respect to the specific lepto-hadronic jet model considered.

\noindent\section{Dark matter-nucleon interaction with a scalar}
\label{app:ScalarMediator}

We now add to the the SM a new scalar mediator $\phi$ with mass
$m_\phi$, instead of a vector mediator.
The interaction Lagrangian coupling $\phi$ to the first-generation quarks $q=u,d$ and the DM candidate $\chi$ is given by
\begin{equation}\label{eq:scalar}
    \mathcal{L}_{\phi} = g_{\chi \phi} \bar{\chi}\chi \phi + g_{q \phi} \bar{q}q \phi \,,
\end{equation}
\begin{figure}[t]
    \centering
    \includegraphics[width=0.5\textwidth]{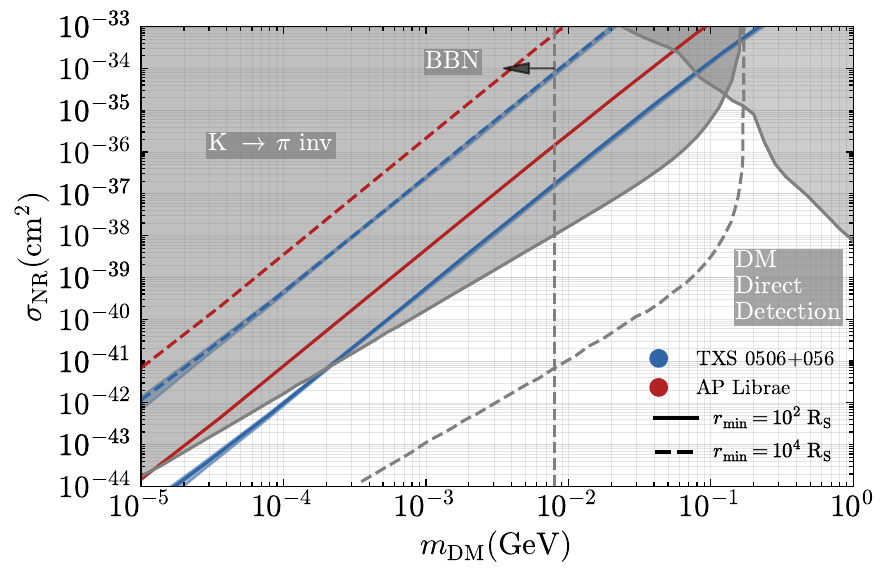}
    \caption{DM parameter space of Eqs~(\ref{eq:scalar}), (\ref{eq:sigmascalar}) for $m_\phi= 5\text{ GeV}$. BBN and DD bounds as in Fig.~\ref{fig:DM}, the grey shaded area on the left is excluded by rare kaon decays constraints~\cite{Cox:2024rew}. For the latter, limits further depend on the coupling of $\phi$ to the top quark $g_{t \phi}$. We report as grey (grey dashed) lines the bounds for $g_{t \phi}=0$ ($g_{t \phi}\neq 0$) corresponding to the gluon-coupled (quark-coupled) cases shown in Fig. 2 of~\cite{Cox:2024rew}.}
    \label{fig:scalarDM}
\end{figure}
We focus on $m_\phi\simeq \mathcal {O} (1 \text{ GeV})$,
as lighter mediators can be produced by meson decays and are strongly constrained~\cite{Batell:2018fqo}. 
In our analysis we consider the isoscalar coupling $g_{u\phi}=g_{d \phi}=g_\phi$. This model, in addition to BBN and direct detection bounds, which are common to the vector case, is also severely constrained by rare kaon decays~\cite{Cox:2024rew}.
In Fig.~\ref{fig:scalarDM} we translate our bounds on the couplings for the scalar mediator case into the more familiar bounds on the non-relativistic cross section, which is given by 
\begin{equation}\label{eq:sigmascalar}
\sigma_\text{NR}=\left(\frac{m_p}{m_u}f_u^p +\frac{m_p}{m_d}f_d^p\right)^2\,\frac{g_{\chi\phi}^2 g_{u\phi}^2}{\pi}\frac{\mu_{\chi p}^2}{m_\phi^4}\,,
\end{equation}
where $m_{u,d}$ are the up and down quark masses, and $[(m_p/m_u)f_u^p +(m_p/m_d)f_d^p]^2\simeq 300$, with $f_q^p \equiv m_q\bra{p}\bar{q}q\ket{p}/(2m_p^2)$ computed at zero momentum transfer.
The latter can be extracted from data and lattice computations: $f_u^p\simeq 1.97 \times 10^{-2}$, $f_d^p\simeq 3.83 \times 10^{-2}$ ~\cite{Hoferichter:2023ptl}.
We find that the scalar mediator scenario cannot account
for the measured neutrino flux by TXS 0506+056 due to the severe
constraints set by rare kaon decays, unless $m_{\DM}\lesssim 0.2\,\text{MeV}$ for BMCI
and the scalar mediator does not couple to the top quark.
See \cite{DeMarchi:2025xag} for a study of the available parameter space in other DM models.

%\bibliography{main.bib}
%\bibliographystyle{JHEP}

\providecommand{\href}[2]{#2}

\end{document}